\documentclass[12pt]{article}

\begin{document}
 \title{The Holographic Ward identity: Examples from 2+1 gravity}
 \author{M\'aximo Ba\~nados  and Rodrigo Caro\\  Departamento de F\'{\i}sica,\\
 P. Universidad Cat\'olica de Chile, Casilla 306, Santiago 22,Chile. \\
 {\tt mbanados@fis.puc.cl, rmcaro@puc.cl}}

 \maketitle

\begin{abstract}
In the AdS/CFT correspondence the boundary Ward identities are encoded in the bulk constraints. 
We study the three-dimensional version of this result using the Chern-Simons formulation of gravity. Due to the metric boundary conditions the conformal identities cannot be derived in a straightforward way from the chiral ones. We pay special attention to this case and find the necessary modifications to the chiral currents in order to find the two Virasoro operators.  The supersymmetric Ward identities are studied as well.  
\end{abstract}

\section{Introduction}

One of the most studied examples of the $AdS/CFT$ correspondence
is the relation between three-dimensional gravity with a negative
cosmological constant and two-dimensional Conformal Field Theory.
Even before the $AdS/CFT$ was discovered, Brown and Henneaux \cite{BH} found
a deep connection between 3d gravity with anti-de Sitter boundary
conditions and the 2d centrally-extended conformal algebra. This
connection was then reformulated in \cite{CHvD} as a relationship
between three dimensional gravity and Liouville theory (and
generalized in \cite{HMS} to supergravity). Within the $AdS/CFT$
correspondence one can understand this result as a relationship
between fluctuations on the $AdS$ background (the Liouville field)
and vacuum expectation values of the 2d $CFT$.

It is unfortunate that  quantum Liouville theory is not fully
understood. There are questions one can ask at the gravity side
which do not yet have an answer within Liouville theory, e.g., the
black hole microstates (see, for example,  \cite{BHE}).
Whether or not Liouville theory encodes the necessary information to answer all
questions related to 3d dimensional quantum gravity is still not
known. See \cite{JT} for a recent discussion on Liouville theory. 

The relationship between 3d gravity and Liouville theory
discovered in \cite{CHvD} is based on the fact that three-dimensional
gravity does not have any local degrees of freedom. Using the relationship between
Chern-Simons and WZW theories \cite{Witten89,MS} together
with the Hamiltonian reduction technics \cite{Liouville}, 
it was shown in \cite{CHvD} that the (local) relevant dynamical 
information is encoded in the Liouville action.  From a
Hamiltonian point of view, the Liouville field describes those
``would-be-trivial" degrees of freedom associated to
diffeomorphisms which preserve the boundary conditions, but have
non-zero generators \cite{RT}.

On the other hand, in the general $AdS/CFT$ correspondence\cite{Maldacena,GKP,Witten98}
the $CFT$ correlators are computed directly from a
classical action on the $AdS$ background. In particular, if one is
interested in correlators involving the energy-momentum tensor,
the needed action is precisely the (renormalised) gravity action
$I_{ren}[g_{ij}]\equiv W[g_{ij}]$ as a function of an {\it
arbitrary} induced boundary metric $g_{ij}$ which acts as a source in
the $CFT$. Performing a near boundary analysis the needed
divergences and counterterms can be identified, and $W[g_{ij}]$
can be constructed explicitely\cite{K}.

The $AdS/CFT$ correspondence establishes also a relationship between the $CFT$ Ward identities 
and RG equations with the bulk equations. The radial (``evolution") equation is 
associated to the $CFT$ renormalization equation, while the
constraints to Ward identities \cite{deBoerV2,Muck,Corley}. This relationship 
becomes particularly nice and clear in three-dimensional Chern-Simons gravity. In fact, it has been known
for a long time \cite{Drinfeld,Polyakov,Verlinde2} that a
two-dimensional flat $SL(2,\Re)$ current encodes the Virasoro Ward
identity, and this result fits quite naturally within $AdS/CFT$.  In this paper we shall put these results in the $AdS/CFT$ context. Just as a single $SL(2,\Re)$ field contains the Virasoro Ward identity, the $SL(2,\Re)\times SL(2,\Re)$ theory contains the conformal identity. However, the construction is slightly more complicated because the boundary conditions couple the left and right gauge fields. We shall see that the $AdS/CFT$ prescription provides a natural reason to decouple both sectors leading to the conformal identities.  We shall also see that the $AdS/CFT$ classification of the gauge field components (leading, first subleading and subleading) provides a natural, and free of ad-hoc assumptions, derivation of these identities from the bulk constraints. 

The paper is organized as follows. After a brief introduction to the Virasoro Ward identity and Polyakov's action, in Section 2 we analyze the bosonic conformal Ward identity, derived from $AdS_3/CFT_2$. In section 3 we consider the $Osp(2,1)\times Osp(2,1)$ theory and find the superconformal Ward identities. Finally, we also consider the extended models in the chiral case.

\section{Bosonic $\mathbf{AdS_3/CFT_2}$}

\subsection{The Virasoro Ward identity}

Let us start by recalling that in bosonic case the generator of energy-momentum tensor correlators, $W[g]$, is given by
Polyakov's non-local action
\begin{equation}\label{PolAc}
W[g] = {c \over 24\pi}\int\!\!\int R \, {1 \over \nabla^2} \, R
\end{equation}
where $c$ is the central charge. The derivation of this functional
from $AdS/CFT$ was studied in \cite{SSol} (see also \cite{BCR}).
In Polyakov's light cone gauge 
\begin{equation}\label{lcg}
ds^2 = dzd\bar z + h(z,\bar z) d\bar z^2.
\end{equation}
the metric has only one independent function $h(z,\bar z)$ and one defines $T = \delta W/\delta h$.
Then, directly from the trace anomaly and conservation of $T^{ij}$ it follows  that $T$
satisfies \cite{Polyakov},
\begin{equation}\label{Ward0}
\bar\partial T =- {c \over 12} \partial^3 h + 2\partial h T + h \partial T
\end{equation}
which is known as the Virasoro Ward identity.  

This relation is contained in a flat $SL(2,\Re)$ 2d gauge field as follows \cite{Drinfeld}.  
Consider in two dimensions the gauge field $A = A_zdz + A_{\bar z}d\bar z$ with the following particular values
\begin{equation}\label{A}
A_z = \left(\begin{array}{cc}
  0 & 1 \\
  T/k & 0 \\
\end{array}\right), \ \ \ \ \ \ \ \ \ \ \ \
A_{\bar z} = \left(\begin{array}{cc}
  \alpha & \mu \\
  \beta & -\alpha \\
\end{array}\right).
\end{equation}
All functions $T,\alpha,\beta,\mu$ are functions of $z$ and $\bar
z.$ The flatness condition $F_{z\bar z}=\partial A_{\bar z} -
\bar\partial A_z + [A_z,A_{\bar z}]=0$ implies
\begin{eqnarray}
  \alpha &=& {1 \over 2} \partial\mu \\
  \beta &=&  {1 \over k}\: \mu T - {1 \over 2} \partial^2\mu \\
  \bar\partial T &=& \mu \partial T + 2\partial\mu\: T - {k \over 2}\partial^3 \mu \label{dT}
\end{eqnarray}
We recognize the third equation (\ref{dT}) as the Virasoro Ward identity with $c=6k$.

Since three-dimensional gravity can be formulated as the difference of $SL(2,\Re)$ Chern-Simons
actions\cite{AW} it is hardly surprising that, within $AdS/CFT$, the Virasoro Ward identities will arise through this mechanism. Actually, the relationship between $SL(2,\Re)$ Chern-Simons theory (from where the constraint $F_{z\bar z}=0$ can be derived) and the 2-dimensional action (\ref{PolAc}) was studied in detail in \cite{Verlinde2}, almost ten years before the $AdS/CFT$ correspondence became available. A similar idea was applied to $W$-gravity in \cite{deBoer}. Our goal in this paper is to put some of the results of \cite{Verlinde2} in the $AdS/CFT$ perspective, clarifying some issues, for example how to deal with the second $SL(2,\Re)$ field, and to extend them to supergravity.    

We shall see that the $AdS/CFT$ correspondence provides a natural interpretation 
on why the above mechanism works. In particular we shall see that $\mu$ enters as the source associated to $T$, $\alpha$ is a subleading term related locally to the source, and $\beta,T$ are non-local, as in (\ref{dT}).  
We shall also extend this mechanism to the non-chiral case, which is not trivial due to the boundary conditions on the metric. 

\subsection{Classical three-dimensional equations and the Conformal Ward identity}

We start by writing the solution to the three-dimensional gravitational equations in the Fefferham-Graham (FG) form. In this paragraph we follow closely \cite{BCR}.  

Let $A$ and $\bar A$ be two $SL(2,\Re)$ fields related to the triad and spin connection in the usual way
\begin{equation}
A = w  + e, \ \ \ \ \ \   \bar A = w-e
\end{equation}
The solution to the equations of motion can be locally written as
\begin{equation}\label{gh}
A = g^{-1}dg,  \ \ \ \ \ \  \bar A = h^{-1}dh
\end{equation}
for arbitrary group elements $g$ and $h$. Since (\ref{gh}) is the
most general local solution to the equations of motion  we should
be able to find the Fefferham-Graham (FG) \cite{FG} expansion for
the metric by manipulating (\ref{gh}). Incidentally, we recall
that, in three dimensions, the FG expansion is finite and has only
three-terms \cite{SSol} (see \cite{B} for a different formulation
of this fact). 

The FG expansion makes use of a choice of coordinates, namely,
$g_{r i}=0$. Our first task is then to write the solutions to the
equations of motion (\ref{gh}) in a way consistent with the FG
expansion, and coordinate choice.  As shown in \cite{BCR} the FG expansion is implemented 
in the Chern-Simons formulation by writting the fields in the form,
\begin{eqnarray}
A(x,r) = e^{-r J_3} \, A(x) \, e^{r J_3} + dr J_3 \label{radial1} \\
\bar A(x,r) = e^{r J_3} \, \bar A(x) \, e^{-r J_3} - dr J_3. \label{radial2}
\end{eqnarray}
where $A(x)=A_i(x)dx^i$ and $\bar A(x)=\bar A_i(x) dx^i$ are both
$SL(2,\Re)-$valued connections defined on the
surface at fixed $r$. The radial dependence is then explicitly 
factor out as a gauge transformation. The fields  
$A(x,r)$ and $\bar A(x,r)$ will satisfy the equations of motion provided 
$A(x)$ and $\bar A(x)$ are both {\it flat}. 

The fields $A(x)$ and $\bar A(x)$ are in the $SL(2,\Re)$ algebra and we can expand them in the basis
$J_+, J_-,J_3$, namely,
\begin{equation}\label{}
A(x) = A^+(x) J_+ + A^-(x) J_- + A^3(x) J_3
\end{equation}
and similarly for the other copy. Replacing these expansions in (\ref{radial1},\ref{radial2}), and using   $e^{-r J_3} J_\pm e^{r J_3} = e^{\mp r} \, J_\pm$, we obtain,
\begin{eqnarray}
A(x,r)  &=& (dr + A^3(x)) J_3 +   e^{-r }\, A^+(x) J_+ + e^{r }\,
          A^-(x) J_-,    \label{A2}\\
\bar A(x,r)  &=&  (-dr +\bar A^3(x)) J_3 +  e^{r }\, \bar A^+(x) J_+
 + e^{-r }\, \bar A^-(x) J_-.
\label{Ab2}
\end{eqnarray}

We now see that the different components of $A(x)$ and $\bar A(x)$
appears multiplied by different powers of $e^r$. In the limit $r\rightarrow\infty$, this expansion
assigns $A^-$ the role of leading term (``source"), $A^3$ the next to leading term,
and $A^+$ as the subleading term, which will be non-local in the source.  The expansion has only three
terms as anticipated previously \cite{SSol}.   A similar analysis holds for
$\bar A$.

Let us now compute the corresponding three-dimensional triad \cite{BCR}
\begin{eqnarray}
e & = & \left( dr  + {1 \over 2}(A^3 - \bar A^3)\right)J_3
    + {1 \over 2} e^r ( A^- J_- -\bar A^+J_+)
    + {1 \over 2} e^{-r} ( A^+ J_+ -\bar A^- J_-).
\label{e}
\end{eqnarray}
Since $A(x)$ and $\bar A(x)$ are forms on the 2-surface, this
vielbein does not have the FG form. In fact the
metric $g_{\mu\nu} = e^a_{\ \mu} e^b_{\ \nu}\eta_{ab}$ will have a piece of the form $dr (A^3_i-\bar A^3_i)dx^i$.
In order to conform with the FG choice of coordinates we need to
impose the extra boundary condition \cite{BCR}
\begin{equation}\label{cond}
A^3(x) = \bar A^3(x).
\end{equation}
This condition turns out to be the key input in order 
to find the Virasoro Ward identities, including the
supersymmetric ones.  In particular, we stress that it
is {\it not} correct to impose $A^3_z=0$ and $\bar A^3_{\bar z}=0$
separately, as one would naively conclude from (\ref{A}). 
Only their difference has to be zero, and this
condition follows in a natural way from the FG choice of
coordinates.

Summarizing, we have two $SL(2,\Re)$ connections $A$ and $\bar A$
defined on the surface at fixed $r.$  They are completely
arbitrary except for the condition (\ref{cond}).  Now, the
different components of $A$ and $\bar A$ play different roles
within the $AdS/CFT$ prescription.  The leading components, $A^-$
and $\bar A^+$, are fixed and represents the sources for the
corresponding $CFT$ operator, analogous to $g_{(0)ij}$ in the
metric FG expansion.  The other components are expressed in terms
of the fixed components $A^-$ and $\bar A^+$ via the equations of
motion. Some of these relations are local, and others non-local.
The 1-point functions will be  related non-locally to $A^-$ and
$\bar A^+$.

The sources $A^-,\bar A^+$ are arbitrary and we can consider
particular values for it, if we only want to turn on some
particular operators.  For example, Polyakov's light cone gauge
(\ref{lcg}) leaves enough information to find one copy of the
Virasoro algebra.  We would like to consider here a more general
situation in which both copies of the Virasoro algebra arise simultaneously.  
To this end, we choose complex coordinates $z,\bar z$ and consider boundary metrics having two Beltrami parameters, 
\begin{equation}\label{P}
ds^2_{(0)} = |d\bar z + \mu(z,\bar z) dz|^2.
\end{equation}
The appropriated bulk gauge field that has as leading term the metric
(\ref{P}), and fulfills (\ref{cond}), is
\begin{eqnarray}
A_{z}(z,\bar z)=\left( \begin{array}{cc} \omega/2 & T'/k \\ 1 & -\omega/2
\end{array} \right) \qquad
\overline{A}_{z}(z,\bar z)=\left(
\begin{array}{cc} \omega/2 & \bar \mu \\ \bar\gamma &
-\omega/2
\end{array} \right) \label{SL2R1}\\
A_{\bar z}(z,\bar z)=\left(
\begin{array}{cc} \bar \omega/2 & \gamma \\ \mu & -\bar \omega/2 \end{array} \right)
\qquad \overline{A}_{\bar z}(z,\bar z)=\left(
\begin{array}{cc} \bar \omega/2 & 1 \\ \bar T'/k & - \bar \omega/2 \end{array}
\right) \label{SL2R2}
\end{eqnarray}
Here all functions depend on $z,\bar z$. $\mu$
and $\bar\mu$ are the sources;  they enter as the leading terms when (\ref{SL2R1},\ref{SL2R2}) are replaced in the  expansion (\ref{radial1},\ref{radial2}). All other components become functions of  $\mu$
and $\bar\mu$ via the equations of motion.

The bulk constraint equations $F=\partial
A_{\overline{z}}-\overline{\partial} A_{z} +
[A_{z},A_{\overline{z}}]=0$ (and similarly for $\bar F=0$) imply the following relations, which can be understood as the non-chiral version of (\ref{dT}), 
\begin{eqnarray}
\omega&=&\bar \omega \bar \mu+\overline{\partial}\bar \mu\label{T1}\\
\bar \omega&=&\omega\mu-\partial\mu \label{T2}\\
\gamma &=& \frac{1}{2}(\partial \bar \omega-\overline{\partial}\omega)+\mu T'/k \label{g}\\
\bar \gamma &=& \frac{1}{2}(\partial \bar
\omega-\overline{\partial}\omega)+
\bar \mu \bar T'/k\label{gb}\\
\overline{\partial}T'/k &=& \partial\gamma+\omega\gamma- \bar \omega T'/k  \label{w1} \\ 
\partial \bar T'/k &=& \overline{\partial}\delta-\bar
\omega \delta+\omega\bar T'/k \label{w2}
\end{eqnarray}
As anticipated previously, $\omega$ and $\bar\omega$ can be expressed locally in terms of the sources $\mu$ and $\bar\mu$ by solving (\ref{T1},\ref{T2}).  Incidentally, we mention that the 1-form $w=\omega dz+\bar \omega d\overline{z}$ turns out to be the spin connection associated to the metric (\ref{P}), 
equations (\ref{T1}-\ref{T2}) are the corresponding torsion free
conditions, and  
\begin{equation}\label{R2}
R=dw
\end{equation}
becomes the two-dimensional curvature.

The functions $\gamma$ and $\bar\gamma$ can also be eliminated algebraically using (\ref{g},\ref{gb}), although they depend on $T'$ and $\bar T'$ which are non-local in the sources.  

Finally, by analogy with (\ref{dT}), one would expect Equations (\ref{w1},\ref{w2}) to give two copies of the Virasoro Ward identity.  To see this, however, one needs to add the following {\it local} pieces to $T'$ and $\bar T'$ \cite{Verlinde2}, 
\begin{eqnarray}
T&=&T'+\frac{k}{2}\big (\frac{1}{2}\omega^{2}+\partial\omega\big), \label{T'} \\
\overline{T}&=&\overline{T}'+\frac{k}{2}\big
(\frac{1}{2}\overline{\omega}^{2}-\overline{\partial}\overline{\omega}\big). \label{Tb'}
\end{eqnarray}
It is direct to see that $T$ and $\bar T$ satisfy the correct Ward identities
\begin{eqnarray}
\overline{\partial}T &=& -\frac{k}{2}\partial^{3}\mu+2(\partial\mu) T
+ \mu \partial T \\
\partial \overline{T}&=& -\frac{k}{2}\overline{\partial}^{3}\overline{\mu}+2(\overline{\partial}\overline{\mu}) \overline{T}
+ \overline{\mu} \overline{\partial} \overline{T}
\end{eqnarray} 

Summarizing, we have shown that the constraint equations of three-dimensional gravity contain the conformal Ward identity.  It remains to prove that the combinations $T$ and $\bar T$ represent the 1-point functions in the presence of the sources $\mu$ and $\bar\mu$. 

\subsection{The action and 1-point vacuum expectation values}

In order to prove that   $T$ and $\bar T$ are the correct 1-point functions we need to consider the renormalised 3d action written as a function of the sources $\mu$ and $\bar\mu$, $W[\mu,\bar\mu]$, and check that 
\begin{equation}\label{dI0}
T=-2\pi{\delta W[\mu,\bar\mu] \over \delta \mu }, \ \ \ \ \ \ \bar
T =-2\pi{\delta W[\mu,\bar\mu] \over \delta \bar\mu }.
\end{equation}
Three-dimensional gravity is given by the difference of two Chern-Simons actions $I[A,\bar A]
= I[A]-I[\bar A]$ with $I[A] = {k \over 4\pi} \int \left( AdA + {2 \over 3} AAA \right)$. We then start by 
looking at the on-shell variation
of $I[A,\bar A]$,
\begin{equation}
\delta I[A,\bar A] = -{k \over 4\pi} \int_{\partial M} (A\delta A
- \bar A \delta \bar A )
\end{equation}
(Here we have omitted the terms that vanishes when the equations of
motion hold.) Replacing here the asymptotic form of the gauge field
(\ref{radial1},\ref{radial2}) and (\ref{g},\ref{gb}) we obtain the {\it finite} result
\begin{equation}\label{dI}
\delta I[A,\bar A] = -{1 \over 2\pi} \int_{\partial M} \left( T'
\delta \mu + \bar T' \delta \bar\mu + \frac{k}{2}\delta R \right)
\end{equation}
where $R = dw$ is the two-dimensional curvature (see (\ref{R2})).   This is almost
what we need. $R$ is a local function of the sources that can be
added (with the opposite sign) to $I[A,\bar A]$, as dictated by the Dirichlet problem
with $\mu$ and $\bar\mu$ fixed.  However, recall that the true 1-point
functions satisfying the correct Ward identities are $T,\bar T$,
defined in (\ref{T'},\ref{Tb'}), and not $T',\bar T'$. However, it
follows that the term which transform $(T',\bar T')$ into $(T,\bar
T)$ in (\ref{dI}) is also a trivial local term.  To see
this consider the following identity
\begin{eqnarray}
   0 &=& \left(\omega \delta \overline{\omega} - \overline{\omega} \delta \omega\right) -
  \left(\omega \delta \overline{\omega} - \overline{\omega} \delta \omega\right) \label{id} \\
   &=& \left( \frac{1}{2}\omega^2+\partial \omega \right)\delta\mu + \left(\frac{1}{2}\overline{\omega}^2
   -  \overline{\partial} \overline{\omega}\right)\delta\overline{\mu}
    +\frac{1}{2}\delta\left(\partial\mu \omega - \overline{\partial}\overline{\mu} \overline{\omega} \right)  \label{id2}
\end{eqnarray}
To go from (\ref{id}) to (\ref{id2}) we have used (\ref{T2}) and (\ref{T1}) in the first and second terms in
(\ref{id}), respectively. We then observe that the first two terms in (\ref{id2}) are exactly the local pieces added to $(T',\bar T')$ in  (\ref{T'},\ref{Tb'}), multiplied by the variations of the sources.  We then find that the sum $\left( \frac{1}{2}\omega^2+\partial \omega \right)\delta\mu + \left(\frac{1}{2}\overline{\omega}^2
   -  \overline{\partial} \overline{\omega}\right)\delta\overline{\mu}$ is equal to the total variation of a local term that we call  $K[\mu,\bar\mu] = \int_{\partial M} \left( \partial\mu w_z - \bar\partial\bar\mu
w_{\bar z} \right)$, and hence trivial.  Note also that $K[\mu,\bar\mu]$ is precisely the interaction term, found in \cite{Verlinde2}, which decouples in the Polyakov action the actions for $\mu$ and
$\bar\mu$. 

In summary, the Dirichlet problem instruct us to consider the improved action (up to divergent counterterms which do not contribute to the 1-point function) $W[\mu,\bar\mu]
:= I[A,\bar A] + \frac{k}{4\pi}\int_{\partial M} \left( R  +
\frac{1}{2}K\right)$ whose variation is given by,
\begin{equation}
\delta W[\mu,\bar\mu] = -\frac{1}{2\pi}\int_{\partial M} \left(
T\delta\mu + \bar T\delta \bar\mu \right),
\end{equation}
in full agreement with (\ref{dI0}), as desired.

\section{Supergravity and the Superconformal Ward identity}

\subsection{Holographic Super Virasoro Ward identities}

We will now generalize the results obtained in the previous
section to the super-symmetric case. The discussion is generalized  
to this case with almost no changes. We then only briefly state the form of the 
currents and their Ward identities.

Three dimensional super-gravity can be recast as the difference of two Chern-Simons
action for the $OSp(1|2)$ group. Let
$\Gamma=A^{a}J_{a}+\chi^{\pm}R_{\pm}$ and $\bar \Gamma=\bar
A^{a}J_{a}+\bar \chi^{\pm}R_{\pm}$ be two $OSp(1|2)$ connections.
The radial coordinate dependence of the connections is given by
(see (\ref{radial1},\ref{radial2})):
\begin{eqnarray}
\Gamma(x,r) &=& (dr + A^{3})J_{3}+e^{-r}A^{+}J_{+}+e^{r}A^{-}J_{-}
+
e^{-r/2}\chi^{+}R_{+}+e^{r/2}\chi^{-}R_{-} \nonumber \\
\bar \Gamma(x,r) &=& (-dr + \bar A^{3})J_{3}+e^{r}\bar
A^{+}J_{+}+e^{-r}\bar A^{-}J_{-} +e^{r/2}\bar
\chi^{+}R_{+}+e^{-r/2}\bar \chi^{-}R_{-} \label{G2}
\end{eqnarray}

Imposing the condition (\ref{cond}) and using (\ref{G2}) to
identify the leading terms, we set the gauge fields to be:
\begin{eqnarray} \label{Osp21}
\Gamma_{z}(x)&=&\left( \begin{array}{ccc} \omega/2 & T'/k &  Q/k\\
1 & -\omega/2 & 0
\\ 0 & Q/k & 0
\end{array} \right) \qquad \bar \Gamma_{z}(x)=\left(
\begin{array}{ccc} \omega/2 & 2\pi \bar \mu &  2\pi \bar \epsilon\\ \delta & -\omega/2 & \theta
\\ -\theta & 2\pi \bar \epsilon & 0
\end{array} \right)\\
\Gamma_{\bar z}(x)&=&\left(
\begin{array}{ccc} \bar \omega/2 & \gamma & \psi \\  2\pi \mu & -\bar \omega/2 &  2\pi \epsilon \\
- 2\pi \epsilon & \psi & 0 \end{array} \right)  \qquad \bar \Gamma
_{\bar z}(x)=\left(
\begin{array}{ccc} \bar \omega/2 & 1 & 0 \\ \bar T'/k & -\bar \omega/2 & \bar Q/k \\
-\bar Q/k & 0 & 0 \end{array} \right)
\end{eqnarray}
where ($\mu,\bar \mu, \epsilon, \bar \epsilon$) are the sources
and the induced 2-d boundary metric is given by (\ref{P}).

Solving for a flat connection, $F=\partial \Gamma_{\bar z}-\bar
\partial \Gamma_{z} + [\Gamma_{z},\Gamma_{\bar z}]=0$ (and $\bar F =0$), we
get the super-Virasoro Ward identities:
\begin{eqnarray}
\frac{1}{2\pi}\bar \partial
T&=&-\frac{k}{2}\partial^{3}\mu+2(\partial\mu) T
+ \mu \partial T +\partial(\epsilon Q)+2(\partial\epsilon)Q\\
\frac{1}{2\pi}\bar \partial Q&=&-k\partial^{2}\epsilon+\epsilon
T+\partial(\mu Q)+\frac{1}{2}\partial\mu Q\\
\frac{1}{2\pi}\partial \bar T&=&-\frac{k}{2}\bar \partial ^{3}\bar
\mu+2(\bar \partial \bar \mu) \overline{T}
+ \bar \mu \bar \partial \bar T+\bar \partial(\bar Q \bar \epsilon)+2\bar Q(\bar \partial \bar \epsilon)\\
\frac{1}{2\pi}\partial \bar Q&=&-k\bar\partial^{2}\bar
\epsilon+\bar \epsilon \bar T +\bar \partial(\bar \mu \bar
Q)+\frac{1}{2}\bar \partial \bar \mu \bar Q \label{S-WardId}.
\end{eqnarray}
Here we have made the same redefinition (\ref{T'},\ref{Tb'}) to go from $T',\bar T'$ to $T$,
$\bar T$.

The functions $T,\overline{T},Q,\overline{Q}$ can be identified with the 2-d CFT
1-point functions, and they can be derived from the
renormalised 3-d super-gravity action written in terms of the
sources.  The action (up to divergent counterterms which do not change the variation) is, again,
\begin{equation}
W[\mu,\bar \mu, \epsilon, \bar \epsilon]=I[\Gamma, \bar
\Gamma]+\frac{k}{4\pi}\int_{\partial M}(R+\frac{1}{2}K[\mu,\bar
\mu]) \label{SuperWAction}
\end{equation}
whose variation becomes
\begin{equation}
\delta W[\mu,\bar \mu, \epsilon, \bar
\epsilon]=-\frac{1}{2\pi}\int_{\partial M}(T\delta \mu +\bar
T\delta \bar \mu+ \delta \epsilon Q + \bar Q\delta \bar \epsilon)
\end{equation}
as desired.

\subsection{Extended Super Virasoro Ward Identities}

We will now obtain the Virasoro Ward identities in the
$\mathcal{N}$-extended super-symmetric case. In this case, we 
work with two copies $Osp(N|2)$ super-group.

The generalization of (\ref{Osp21}) allows us to impose on both
connections the following conditions (we will work here only with the chiral
sources $\mu,\epsilon_{\rho}$ and set $\bar \mu=\bar
\epsilon_{\rho}=0$):
\begin{eqnarray}
\Gamma_{x}&=&\frac{1}{2}\omega J_{3}+\frac{1}{k}T J_{+}+J_{-}+\frac{1}{k}Q_{\rho}R^{\rho}_{+}+E^{a}B_{a}\\
\Gamma_{x}&=&\frac{1}{2}\bar \omega  J_{3}+\gamma J_{+}+\mu
J_{-}+\psi_{\rho}R^{\rho}_{+}+\epsilon_{\rho}R^{\rho}_{-}+\zeta^{a}B_{a}\\
\bar \Gamma_{z}&=&\frac{1}{2}\omega J_{3}+\delta J_{-}+\theta_{\rho}R^{\rho}_{-}\\
\bar \Gamma_{\bar z}&=&\frac{1}{2}\bar \omega J_{3}+ J_{+}+\xi
J_{-}+\chi_{\rho}R^{\rho}_{-} \label{OspN2}
\end{eqnarray}
where we have imposed the condition (\ref{cond}). This condition
does not affect the $SO(N)$ generators, $B_{a}$, which are
anti-symmetric matrices.

Solving for a flat connection, we can find the
$\mathcal{N}$-extended Super Virasoro-Ward identities:
\begin{eqnarray}
\bar\partial T&=&-\frac{1}{2}\partial^{3}\mu+\partial(\mu
T)+(\partial\mu) T
-\eta^{\rho\sigma}\big(\frac{1}{2}\partial(Q_{\rho}\epsilon_{\sigma})-\partial
\epsilon_{\rho} Q_{\sigma}\big)-
(\lambda_{a})_{\phantom{\sigma}\rho}^{\sigma}E^{a}\epsilon^{\rho}Q_{\sigma}\\
\bar \partial Q_{\rho}&=&-\partial^{2}
\epsilon_{\rho}+\partial(\mu Q_{\rho})+ \frac{1}{2}(\partial\mu)
Q_{m} + \epsilon_{\rho} T+(\lambda_{a})_{\sigma\rho}\big(\partial(\epsilon^{\sigma}E^{a})+ E^{a}\partial \epsilon^{\sigma}\big)\\
 &\phantom{=}&+ \zeta^{a}Q^{\sigma}(\lambda_{a})_{\sigma\rho}
 - \mu(\lambda_{a})_{\sigma\rho} E^{a}Q^{\sigma} +(\lambda_{a})_{\sigma\rho}(\lambda_{b})_{\phantom{\sigma}\varrho}^{\sigma}\epsilon^{\varrho}E^{a}E^{b}\\
\bar \partial E^{a}&=&\partial
\zeta^{a}+f^{a}_{\phantom{a}bc}E^{b}\zeta^{c}+\kappa
Q^{\sigma}\epsilon^{\rho}(\lambda^{a})_{\sigma\rho}
\end{eqnarray}

\section{Concluding remarks}

We have studied in this paper the relationship between supergravity bulk contraints and CFT boundary Ward identities, as predicted by the $AdS/CFT$ correspondence. We have concentrated in the $AdS_3/CFT_2$ case and have made use of the Chern-Simons formulation for the 3-dimensional bulk supergravity theory. The bulk constraints are then represented by the rather simple equation $F_{z\bar z}=0$ where $F$ is the associated Yang-Mills curvature. The Ward identities are shown to be encoded in this equation, provided the boundary conditions for the gauge field (AdS asymptotics in the Feffermam-Graham sense) are properly incorporated. 

We have only considered the saddle point evaluation for the bulk partition function. It would be interesting to go further and compute the full path integral, including also other insertions. Since Chern-Simons theory has no local degrees of freedom this seems feasible. On the other hand, the local group is non-compact, $Osp(2,1)\times Osp(2,1)$, and hence the calculation will be challenging. 

Finally, it would also be interesting to study the generalization of this approach to the $AdS_5/CFT_4$ correspondence when the bulk action is taken to be 5-dimensional Chern-Simons supergravity \cite{Chamseddine}. The holographic energy-momentum tensor for this theory has recently been computed in \cite{BST}, following the cohomological methods introduced in \cite{ST}. The associated Ward identities would probe other aspects of the corresponding dual CFT.

\section{ Acknowledgments}

We would like to thank Marc Henneaux, Kostas Skenderis, Adam Schwimmer and Stefan Theisen  for useful discussions on $AdS/CFT$.  MB was partially supported by Fondecyt, CHILE, Grant 1020832.

\section{Conventions}

The $OSp(1|2)$ group has five generators, three bosonic (which
close under $SL(2,\Re)$ Lie algebra) and two fermionic ones. We
will use the following representation for the generators:
\begin{eqnarray}
J_{+}&=& \left( \begin{array}{ccc} 0 & 1 & 0 \\ 0 & 0 & 0 \\ 0 & 0
& 0 \end{array} \right) \qquad J_{-}= \left(\begin{array}{ccc} 0 &
0 & 0 \\ 1 & 0 & 0 \\ 0 & 0 & 0 \end{array} \right) \qquad J_{3}=
\frac{1}{2} \left( \begin{array}{ccc} 1 & 0 & 0 \\ 0 & -1 & 0 \\ 0
& 0 & 0 \end{array} \right) \\
R_{+}&=& \frac{1}{\sqrt{2}}\left( \begin{array}{ccc} 0 & 0 & 1 \\ 0 & 0 & 0 \\
0 & 1 & 0 \end{array} \right) \qquad R_{-}= \frac{1}{\sqrt{2}}
\left(\begin{array}{ccc} 0 & 0 & 0 \\ 0 & 0 & 1 \\ -1 & 0 & 0
\end{array} \right) \label{OSp12.Rep}
\end{eqnarray}
The generators satisfy the following:
\begin{eqnarray}
STr(J_{+}J_{-})&=&1 \qquad STr(J_{3}J_{3})=\frac{1}{2}\\
STr(R_{\pm}R_{\mp})&=&\pm 1
\end{eqnarray}

The $Osp(1|2)$ group can the generalized to the
$\mathcal{N}$-extended super-symmetric case, to get the
$Osp(\textsl{N}|2,R)$ group, whose super-Lie algebra is
the following:

$\bullet$ The three bosonic generators satisfy (which is just the
$SL(2,R)$ Lie algebra):
\begin{equation}
[J_{3},J_{\pm}]=\pm J_{\pm} \qquad [J_{+},J_{-}]=2J_{3}
\end{equation}
$\bullet$ The internal symmetry group generators (which is the
$SO(\textsl{N})$ algebra) satisfy:
\begin{equation}
[B_{a},B_{b}]=f_{abc}B^{c}
\end{equation}
where $a,b,c = 0\ldots N$.

$\bullet$ The fermionic generators satisfy:
\begin{equation}
\{R^{\rho}_{\pm},R^{\sigma}_{\pm}\}=\pm\eta^{\rho\sigma}J_{\pm}
\qquad
\{R^{\rho}_{\pm},R^{\sigma}_{\mp}\}=-\eta^{\rho\sigma}J_{3}\pm\kappa(\lambda^{a})^{\rho\sigma}B_{a}
\end{equation}
where $\lambda^{a}=(\lambda^{a})^{\alpha}_{\beta}$ is the basis of
the representation and satisfy the relation
$[\lambda^{a},\lambda^{a}]=f^{abc}\lambda_{c}$, and $\kappa$ is a
normalization constant.

$\bullet$ All the generators are related by:
\begin{eqnarray}
[J_{3},R^{\rho}_{\pm}]&=&\pm\frac{1}{2}R^{\rho}_{\pm} \qquad
[J_{\pm},R^{\rho}_{\pm}]=0\qquad
[J_{\pm},R^{\rho}_{\mp}]=R^{\rho}_{\pm}\\
 \qquad
[B_{a},R^{\rho}_{\pm}]&=&-(\lambda_{a})_{\phantom{\rho}\sigma}^{\rho}R^{\sigma}_{\pm}
\end{eqnarray}



\begin{thebibliography}{10}

\bibitem{BH}
J.~D.~Brown and M.~Henneaux,
Commun.\ Math.\ Phys.\  {\bf 104}, 207 (1986).

\bibitem{CHvD}
O.~Coussaert, M.~Henneaux and P.~van Driel,
Class.\ Quant.\ Grav.\  {\bf 12}, 2961 (1995)
[arXiv:gr-qc/9506019].

\bibitem{HMS}
M.~Henneaux, L.~Maoz and A.~Schwimmer,
Annals Phys.\  {\bf 282}, 31 (2000)
[arXiv:hep-th/9910013].

\bibitem{BHE}
S.~Carlip,
Class.\ Quant.\ Grav.\  {\bf 15}, 3609 (1998)
[arXiv:hep-th/9806026].
Y.~j.~Chen,
Class.\ Quant.\ Grav.\  {\bf 21}, 1153 (2004)
[arXiv:hep-th/0310234].
O.~Aharony, S.~S.~Gubser, J.~M.~Maldacena, H.~Ooguri and Y.~Oz,
Phys.\ Rept.\  {\bf 323}, 183 (2000)
[arXiv:hep-th/9905111].
E.~J.~Martinec,
arXiv:hep-th/9809021.
J.~Troost and A.~Tsuchiya,
JHEP {\bf 0306}, 029 (2003)
[arXiv:hep-th/0304211].

\bibitem{JT}
J.~Teschner,
Class.\ Quant.\ Grav.\  {\bf 18}, R153 (2001)
[arXiv:hep-th/0104158].

\bibitem{Witten89}
E.~Witten,
Commun.\ Math.\ Phys.\  {\bf 121}, 351 (1989).

\bibitem{MS}
G.~W.~Moore and N.~Seiberg,
Phys.\ Lett.\ B {\bf 220}, 422 (1989).
S.~Elitzur, G.~W.~Moore, A.~Schwimmer and N.~Seiberg,
Nucl.\ Phys.\ B {\bf 326}, 108 (1989).

\bibitem{Liouville}
A.~Alekseev and S.~L.~Shatashvili,
Nucl.\ Phys.\ B {\bf 323}, 719 (1989).
P.~Forgacs, A.~Wipf, J.~Balog, L.~Feher and L.~O'Raifeartaigh,
Phys.\ Lett.\ B {\bf 227}, 214 (1989).

\bibitem{RT}
T.~Regge and C.~Teitelboim,
Annals Phys.\  {\bf 88}, 286 (1974).

\bibitem{Maldacena}
J.~M.~Maldacena,
Adv.\ Theor.\ Math.\ Phys.\  {\bf 2}, 231 (1998)
[Int.\ J.\ Theor.\ Phys.\  {\bf 38}, 1113 (1999)]
[arXiv:hep-th/9711200].

\bibitem{GKP}
S.~S.~Gubser, I.~R.~Klebanov and A.~M.~Polyakov,
Phys.\ Lett.\ B {\bf 428}, 105 (1998)
[arXiv:hep-th/9802109].

\bibitem{Witten98}
E.~Witten,
Adv.\ Theor.\ Math.\ Phys.\  {\bf 2}, 253 (1998)
[arXiv:hep-th/9802150].

\bibitem{K}
M.~Henningson and K.~Skenderis,
JHEP {\bf 9807}, 023 (1998)
[arXiv:hep-th/9806087].
S.~de Haro, S.~N.~Solodukhin and K.~Skenderis,
Commun.\ Math.\ Phys.\  {\bf 217}, 595 (2001)
[arXiv:hep-th/0002230].
M.~Bianchi, D.~Z.~Freedman and K.~Skenderis,
Nucl.\ Phys.\ B {\bf 631}, 159 (2002)
[arXiv:hep-th/0112119].
M.~Bianchi, D.~Z.~Freedman and K.~Skenderis,
JHEP {\bf 0108}, 041 (2001)
[arXiv:hep-th/0105276].
I.~Papadimitriou and K.~Skenderis,
arXiv:hep-th/0407071.


\bibitem{deBoerV2}
J.~de Boer, E.~Verlinde and H.~Verlinde,
JHEP {\bf 0008}, 003 (2000)
[arXiv:hep-th/9912012].

\bibitem{Muck}
J.~Kalkkinen, D.~Martelli and W.~Muck,
JHEP {\bf 0104}, 036 (2001)
[arXiv:hep-th/0103111].

\bibitem{Corley}
S.~Corley,
Phys.\ Lett.\ B {\bf 484}, 141 (2000)
[arXiv:hep-th/0004030].

\bibitem{Drinfeld}
V.~G.~Drinfeld and V.~V.~Sokolov,
J.\ Sov.\ Math.\  {\bf 30} (1984) 1975.

\bibitem{Polyakov}
A.~M.~Polyakov,
Mod.\ Phys.\ Lett.\ A {\bf 2}, 893 (1987).

\bibitem{Verlinde2}
H.~Verlinde,
Nucl.\ Phys.\ B {\bf 337}, 652 (1990);
H.~Verlinde and E.~Verlinde,
PUPT-89-1149
{\it Based on lectures given at the Trieste Spring School, Apr 1989, and at the Ecole Normale Superieure in Paris, Jun 1989}

\bibitem{SSol}
K.~Skenderis and S.~N.~Solodukhin,
Phys.\ Lett.\ B {\bf 472}, 316 (2000)
[arXiv:hep-th/9910023].

\bibitem{BCR}
M.~Banados, O.~Chandia and A.~Ritz,
Phys.\ Rev.\ D {\bf 65}, 126008 (2002)
[arXiv:hep-th/0203021].

\bibitem{AW}
A.~Achucarro and P.~K.~Townsend,
Phys.\ Lett.\ B {\bf 180}, 89 (1986).
E.~Witten,
Nucl.\ Phys.\ B {\bf 311}, 46 (1988).

\bibitem{deBoer}
J.~de Boer and J.~Goeree,
Nucl.\ Phys.\ B {\bf 381}, 329 (1992)
[arXiv:hep-th/9112060].
J.~De Boer and J.~Goeree,
Nucl.\ Phys.\ B {\bf 401}, 369 (1993)
[arXiv:hep-th/9206098].

\bibitem{B}
M.~Banados,
arXiv:hep-th/9901148.

\bibitem{FG}
C. Fefferman and C. R. Graham, "Conformal Invariants," Elie Cartan et les
Math´ematiques d'aujourdhui" (Asterisque, 1985), 95.

\bibitem{Chamseddine}
A.~H.~Chamseddine,
Nucl.\ Phys.\ B {\bf 346}, 213 (1990).

\bibitem{BST}
M.~Banados, A.~Schwimmer and S.~Theisen,
JHEP {\bf 0405}, 039 (2004)
[arXiv:hep-th/0404245].

\bibitem{ST}
C.~Imbimbo, A.~Schwimmer, S.~Theisen and S.~Yankielowicz,
Class.\ Quant.\ Grav.\  {\bf 17}, 1129 (2000)
[arXiv:hep-th/9910267].
A.~Schwimmer and S.~Theisen,
JHEP {\bf 0008}, 032 (2000)
[arXiv:hep-th/0008082].
A.~Schwimmer and S.~Theisen,
JHEP {\bf 0310}, 001 (2003)
[arXiv:hep-th/0309064].









\end{thebibliography}
 \end{document}